\documentclass[aps,prb,twocolumn,superscriptaddress,nofootinbib]{revtex4-2}

\usepackage{graphicx}
\usepackage{amsmath,amssymb}
\usepackage{xcolor}
\usepackage{hyperref}
\hypersetup{colorlinks=true, linkcolor=blue!60!black, citecolor=blue!60!black,
            urlcolor=blue!60!black}

\newcommand{\I}{\mathcal{I}}

\begin{document}

\title{Demons on a Budget: Adaptive Measurement Placement\\
at the Entanglement Phase Transition}

\author{Rohan Pandey}
\email{rpande.1729@gmail.com}
\affiliation{Independent researcher}

\date{\today}

\begin{abstract}
Monitored quantum circuits exhibit a measurement-induced phase transition between volume-law and area-law entanglement as a function of the measurement rate $p$. Prior work places measurements at random locations and treats the rate as the control parameter. We instead fix the measurement budget and vary the placement process, comparing random placement against hand-designed and learned policies in brickwork random Clifford circuits at matched budget. First, placement geometry matters more than placement information. A deterministic contiguous sweep cuts the half-cut entropy by a factor of 3.4 relative to random placement, while equal-coverage unstructured placement and a greedy policy with full state access do far worse. The effect is carried by spatial order alone: measuring the $k$ least recently measured sites gives $4.14 \pm 0.06$ bits with random tie-breaking and $1.29 \pm 0.04$ bits with position-ordered tie-breaking. Second, the sweep eliminates the transition rather than shifting it. Tripartite mutual information crossings recede as $p^* \propto 1/L$, the steady-state entropy saturates at an $L$-independent ceiling near $0.46/p$, and data for $64 \le L \le 512$ collapse onto the form $S = p^{-1} f(pL)$ predicted by a ballistic regrowth argument. Third, in stabilizer dynamics every outcome is deterministic or a fair coin flip, so the record's Shannon entropy is exactly countable; the sweep dominates the entropy-versus-record-cost frontier while paying the same roughly one bit per measurement as random placement. Policies trained by cross-entropy and proximal policy optimization do not find the sweep: score-based policies parameterize which sites to measure, not the order in which degenerate scores are resolved, and the effect lives in that order. The phase diagram of monitored dynamics is a property of the placement process, not only of the measurement rate.
\end{abstract}

\maketitle

\section{Introduction}
\label{sec:intro}

Local measurements compete with entangling unitary dynamics. Below a
critical measurement rate $p_c$ a monitored many-body system retains
volume-law entanglement; above it the state collapses to area
law~\cite{Li2018,Skinner2019,Chan2019}. This measurement-induced phase
transition (MIPT) is a central organizing concept for monitored quantum
matter~\cite{Fisher2023}. Its critical properties have been mapped out for
random Clifford circuits~\cite{Gullans2020,Zabalo2020} and its signatures
have been observed experimentally~\cite{Google2023,Feng2026}.

In nearly all of this literature the measured locations are drawn
uniformly at random. Existing departures from uniform randomness are
static in time. Quasiperiodic measurement sublattices~\cite{Shkolnik2023}
and quenched rate disorder~\cite{Zabalo2023} modify the critical point and
its universality class but preserve the transition. A deterministic
spatial gradient of the rate realizes the transition as coexisting phases
in a single chain~\cite{Li2026space}. Adaptive circuits condition unitary
feedback on measurement outcomes and produce absorbing-state
physics~\cite{Iadecola2023,Sierant2023,ODea2024,Buchhold2022,Piroli2023},
but the measured locations remain random. Closest to our setting,
reinforcement learning has been used to find small sets of projective
measurements that disentangle the output of shallow random circuits; the
structure found there is temporal, a concentration of measurements in late
layers at system sizes of order ten qubits, and the steady-state phase
diagram is not addressed~\cite{Bao2024}. In none of this work is the
spatial placement process itself the designed object.

We take placement as the resource. We fix the expected number of
measurements per layer and ask how much the geometry and the information
content of placement decisions matter, and what they cost. We model the
placing agent as a Maxwell demon that reads part of the classical
measurement record and decides where to look next. In stabilizer dynamics
the demon's ledger can be kept exactly, because each outcome carries
either zero or exactly one bit of Shannon entropy.

The paper is organized as follows. Section~\ref{sec:background} reviews
the quantities used throughout: entanglement entropy and its scaling laws,
monitored circuits, the stabilizer formalism, the tripartite mutual
information, Landauer's principle, and the policy optimization methods.
Section~\ref{sec:model} defines the model, the placement policies, and the
exact record-entropy ledger. Section~\ref{sec:geometry} contains the main
result: at matched budget a deterministic sweeping placement eliminates
the volume-law phase, with finite-size crossings receding as $1/L$ and a
single-parameter scaling collapse across $L = 64$ to $512$, and neither
coverage nor state information reproduces the effect.
Section~\ref{sec:ledger} accounts for all policies on the information
ledger. Section~\ref{sec:learned} shows that policies trained by
reinforcement learning fail to discover the sweep and identifies what
blocks them. Section~\ref{sec:discussion} discusses implications and open
questions. Appendices~\ref{app:numerics} and~\ref{app:learning} give
numerical and training details.

\section{Background}
\label{sec:background}

\subsection{Entanglement entropy and its scaling}
For a pure state $|\psi\rangle$ of a chain partitioned into a region $A$
and its complement $B$, the entanglement entropy of $A$ is the von
Neumann entropy of the reduced density matrix,
\begin{equation}
S_A = -\mathrm{Tr}\,(\rho_A \log_2 \rho_A),
\qquad
\rho_A = \mathrm{Tr}_B\, |\psi\rangle\langle\psi|,
\label{eq:vn}
\end{equation}
measured here in bits. Its dependence on the subsystem size distinguishes
phases of dynamics. In a volume-law phase $S_A$ grows proportionally to
the number of sites in $A$, as it does in the steady state of generic
unitary dynamics. In an area-law phase $S_A$ saturates to a constant
independent of subsystem size. In one dimension the boundary of a
contiguous region is a pair of points, so area law means $S_A = O(1)$.

\subsection{Monitored circuits and the MIPT}
A monitored circuit alternates entangling unitary gates with local
projective measurements applied at rate $p$. Each sequence of measurement
outcomes defines a quantum trajectory, and observables are averaged over
trajectories and circuit realizations. Unitary gates spread entanglement
ballistically while measurements remove it locally, and the competition
produces the measurement-induced phase transition: a volume-law phase for
$p < p_c$ and an area-law phase for
$p > p_c$~\cite{Li2018,Skinner2019,Chan2019}. For the brickwork random
Clifford circuit studied here, $p_c \approx 0.16$~\cite{Zabalo2020}.

\subsection{Stabilizer states and Clifford circuits}
A stabilizer state on $n$ qubits is the unique joint $+1$ eigenstate of
an abelian group $S$ of $n$ independent commuting Pauli operators, its
stabilizer group. Clifford gates map Pauli operators to Pauli operators
and therefore map stabilizer states to stabilizer states, which allows
classical simulation in polynomial time~\cite{AaronsonGottesman2004}.
Writing each of the $n$ generators as a binary vector of $X$ and $Z$
components yields the generator matrix $G$ with $n$ rows and $2n$
columns. Entanglement entropies are then exact rank
computations~\cite{Fattal2004}:
\begin{equation}
S_A = \mathrm{rank}_{\mathrm{GF}(2)}\!\left(G|_A\right) - |A|,
\label{eq:rank}
\end{equation}
where $G|_A$ keeps the columns supported on $A$ and arithmetic is
modulo 2. All entropies in this work are computed from
Eq.~\eqref{eq:rank}, with no sampling error beyond the trajectory
ensemble.

\subsection{Tripartite mutual information}
Locating $p_c$ from $S_A$ directly is contaminated by boundary
contributions. The standard diagnostic subtracts them. For contiguous
quarters $A$, $B$, $C$, $D$ of the chain, the tripartite mutual
information is
\begin{equation}
\I_3 = S_A + S_B + S_C - S_{AB} - S_{AC} - S_{BC} + S_{ABC}.
\label{eq:i3}
\end{equation}
Boundary-law terms cancel in this combination, so $\I_3 \to 0$ in the
area-law phase, while in the volume-law phase $\I_3$ is negative and
extensive. Near a conventional critical point $\I_3$ obeys the
finite-size scaling ansatz
\begin{equation}
\I_3(p, L) = \Phi\!\left((p - p_c)\, L^{1/\nu}\right),
\label{eq:fss}
\end{equation}
with $\nu$ the correlation-length exponent, so curves for different $L$
cross at $p = p_c$ and the crossings of successive size pairs converge to
the critical point~\cite{Gullans2020,Zabalo2020}. A transition exists
only if these crossings converge to a finite $p_c$. Crossings that recede
toward zero with system size signal the absence of a critical point. The
behavior of the crossings is the key diagnostic in
Sec.~\ref{sec:geometry}.

\subsection{Maxwell's demon and Landauer's principle}
A Maxwell demon is an agent that uses measurement information to reduce
the entropy of a system. The resolution of the associated paradox is
bookkeeping: the demon must record outcomes, and by Landauer's principle
erasing one bit of record costs at least $k_B T \ln 2$ of dissipated
heat~\cite{Landauer1961,Bennett1982}. A consistent account of any
measurement-based protocol therefore weighs the entropy it removes
against the record entropy it writes. Section~\ref{sec:model} shows that
in stabilizer dynamics this record entropy is exactly countable, and
Sec.~\ref{sec:ledger} uses it to compare placement policies.

\subsection{Policy optimization}
A placement policy is a map from an observation to a choice of $k$
measurement sites, and an episode is one circuit run scored by a reward,
here the negative final half-cut entropy. We use two standard optimizers.
The cross-entropy method (CEM) is gradient free: it samples policy
parameters from a Gaussian, keeps the best-scoring fraction, and refits
the Gaussian to the elite set. Proximal policy optimization (PPO) is a
policy-gradient method that ascends the reward while clipping each update
to remain close to the current policy, which stabilizes training. Both
optimize the same objective at matched measurement budget. Details appear
in Appendix~\ref{app:learning}.

\section{Model and methods}
\label{sec:model}

\subsection{Monitored Clifford circuits at fixed budget}
We consider a chain of $L$ qubits with periodic boundary conditions. The
chain evolves under brickwork layers of independent uniformly random
two-qubit Clifford gates, interleaved with projective single-site $Z$
measurements. In the standard model each site is measured independently
with probability $p$ per layer. We compare placement processes at matched
budget $pL$ per layer:

\begin{enumerate}
\item[(i)] \emph{Random placement}, the budget-matched analogue of the
standard rate-$p$ model.

\item[(ii)] \emph{The sweep}, which measures a contiguous block that
advances around the ring,
\begin{equation}
\begin{aligned}
A_t &= \{x_t,\, x_t + 1,\, \ldots,\, x_t + m_t - 1\} \bmod L, \\
x_{t+1} &= x_t + m_t,
\end{aligned}
\label{eq:sweep}
\end{equation}
with $m_t \sim \mathrm{Binomial}(L, p)$ drawn independently each layer, so
that $\mathbb{E}\,|A_t| = pL$ matches the random-placement budget exactly
while the rate $p$ remains continuously tunable.

\item[(iii)] \emph{Shuffled coverage}, which visits every site exactly
once per coverage period of $1/p$ layers, as the sweep does, but in a
random order redrawn each period.

\item[(iv)] \emph{Stalest placement}, which measures the $k$ sites least
recently measured. This rule reads only the classical record. It is
degenerate: for the first $1/p$ layers every unmeasured site carries the
same staleness, so the rule fixes the set of sites but not their order. We
resolve the degeneracy two ways, at random and in position order, and
treat the two as separate policies.

\item[(v)] \emph{Oracle greedy} placement at the sites of largest local
entanglement, which uses quantum information unavailable to a physical
observer.

\item[(vi)] \emph{Learned policies} restricted to classical record
information (Sec.~\ref{sec:learned}).
\end{enumerate}

Stalest placement with position-ordered tie-breaking and the sweep produce
identical trajectories: once the first coverage period lays down a
contiguous block, staleness alone advances the block thereafter. We
therefore report them as one policy where the distinction does not matter,
and as two only in Sec.~\ref{sec:order}, where the tie-break is the
variable under study.

Entanglement entropies are computed exactly from Eq.~\eqref{eq:rank}, and
simulations use \texttt{stim}~\cite{AaronsonGottesman2004,Gidney2021}. We
use circuit depth $T = 4L$ and locate transitions with the $\I_3$
crossings of Eq.~\eqref{eq:i3}. Details are given in
Appendix~\ref{app:numerics}.

\subsection{An exact information ledger}
The following standard fact makes the demon's ledger exact in stabilizer
dynamics.

\emph{Lemma (bit dichotomy).} Let $|\psi\rangle$ be a stabilizer state on
$n$ qubits with stabilizer group $S$, and let a projective measurement of
$Z_q$ be performed. Either $\pm Z_q \in S$, in which case the outcome is
deterministic and carries zero Shannon entropy, or $Z_q$ anticommutes with
some element of $S$, in which case the two outcomes occur with probability
$1/2$ each and the outcome carries exactly one bit.

\emph{Proof.} If $\pm Z_q \in S$ the state is an eigenstate of $Z_q$ and
the outcome is fixed. Otherwise choose $g \in S$ with $\{g, Z_q\} = 0$.
Then
$\langle \psi | Z_q | \psi \rangle
 = \langle \psi | g^\dagger Z_q g | \psi \rangle
 = - \langle \psi | Z_q | \psi \rangle = 0$,
so the outcomes are unbiased~\cite{AaronsonGottesman2004,Fattal2004}.
\hfill$\square$

The Shannon entropy of the measurement record in a trajectory therefore
equals the number of nondeterministic outcomes, which we count exactly
during simulation via a pre-measurement determinism check. This number is
the minimum memory a demon implementing the policy must write. By
Landauer's principle, $k_B T \ln 2$ times this number bounds the heat cost
of erasing the record~\cite{Landauer1961,Bennett1982}.

\section{Placement geometry eliminates the transition}
\label{sec:geometry}

\begin{figure}
\includegraphics[width=\columnwidth]{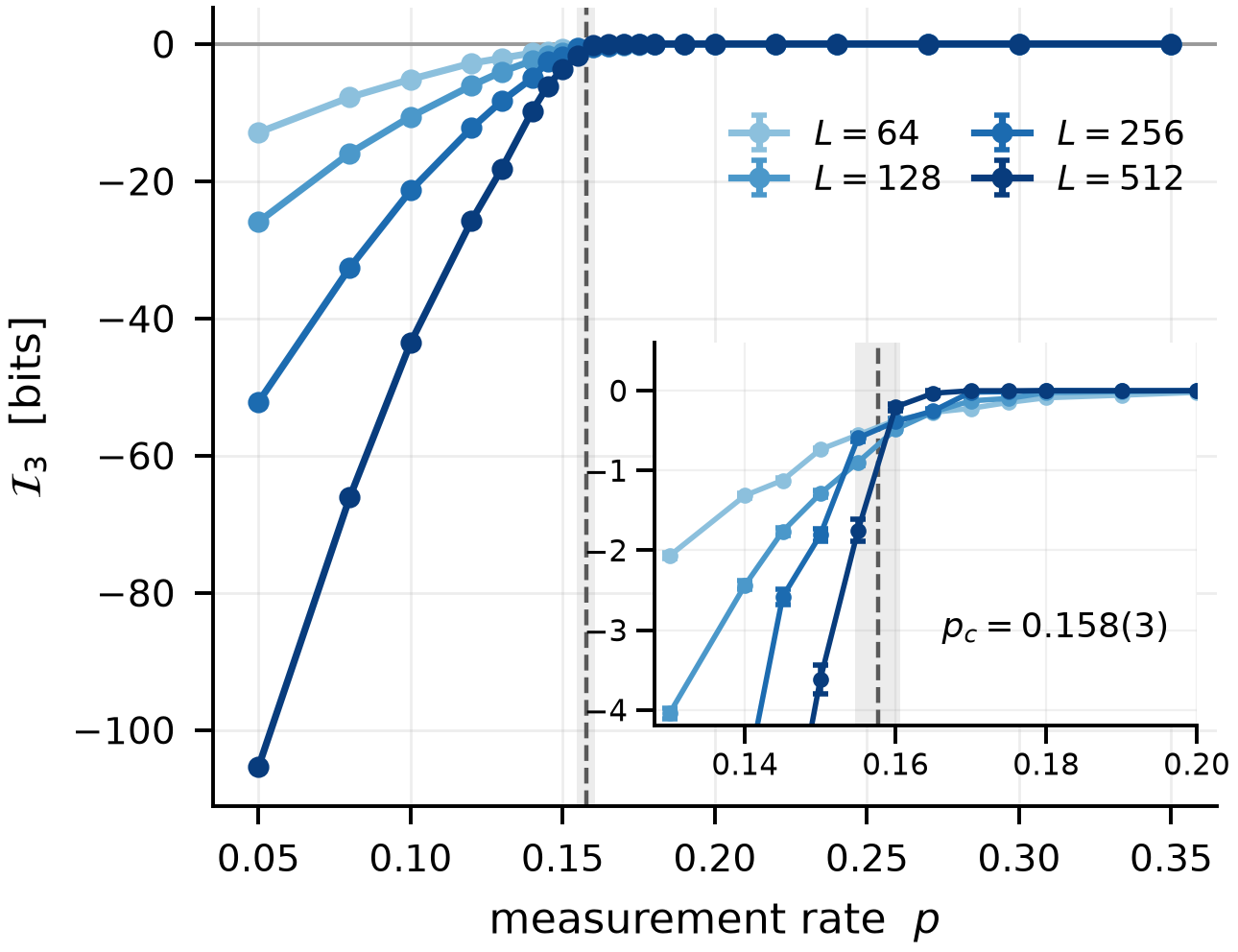}
\caption{Random placement baseline. Tripartite mutual information versus
measurement rate for $L = 64$--$512$ at depth $T = 4L$. Inset: the
crossing region. The dashed line and shaded band give the mean and spread
of the four locally significant pairwise crossings,
$p_c = 0.158 \pm 0.003$, consistent with the established value
$p_c \approx 0.16$ for this model~\cite{Zabalo2020}.}
\label{fig:baseline}
\end{figure}

\begin{figure}
\includegraphics[width=\columnwidth]{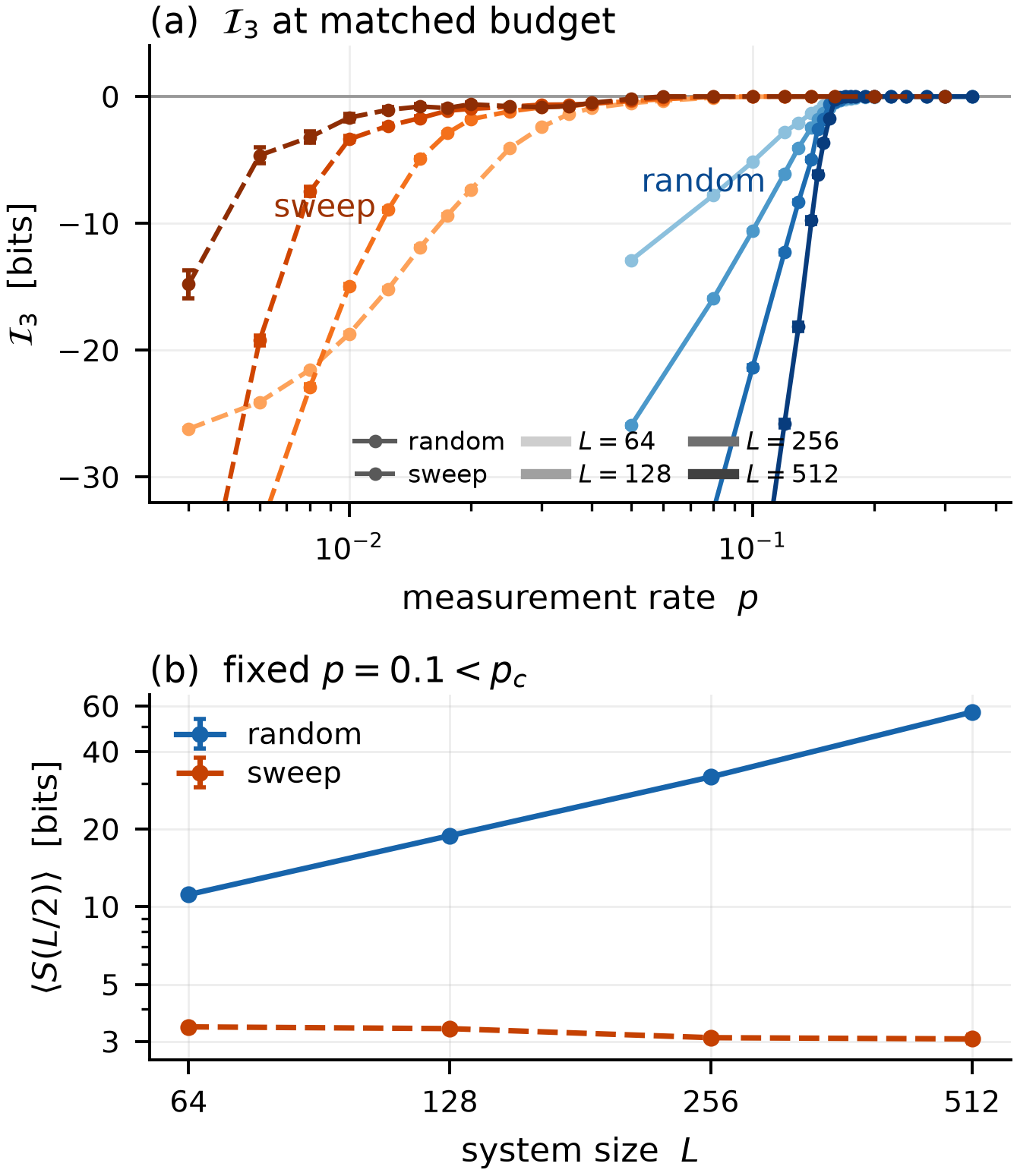}
\caption{Identical budget, different geometry. (a) Sweep placement
(dashed, orange) drives $\I_3$ to zero one to two decades in $p$ below the
random-placement transition (solid, blue); shade encodes $L$. Under random
placement the curves steepen with $L$ and cross at a fixed rate; under the
sweep they separate and recede. (b) At fixed $p = 0.1 < p_c$ the
random-placement entropy grows linearly in $L$ while the sweep entropy is
flat.}
\label{fig:phaseshift}
\end{figure}

\subsection{Baseline}
With random placement, the locally significant pairwise $\I_3$ crossings
among $L = 64$--$512$ cluster at $p_c = 0.158 \pm 0.003$, and the
largest-size pair $(256, 512)$ gives $p_c = 0.159$
(Fig.~\ref{fig:baseline}). Both agree with established results for this
model~\cite{Zabalo2020}. This validates the pipeline at the scales used
below.

\subsection{The sweep}
At matched budget the sweep behaves qualitatively differently
(Fig.~\ref{fig:phaseshift}). Its $\I_3$ crossings do not converge to a
finite rate. Only two pairs cross significantly, and they recede with
size: the $(64,128)$ pair crosses at $p^\ast = 0.0085$ and the $(64,256)$
pair at $p^\ast = 0.0056$. Both satisfy
$p^\ast \sqrt{L_1 L_2} \approx 0.75$, consistent with
$p^\ast \propto 1/L$ at fixed $p^\ast L$. For every pair involving
$L = 256$ or $512$ the curves are statistically indistinguishable over a
broad window in which $|\I_3| \lesssim 1$, and no significant crossing
exists on our grid, consistent with the crossing having receded below
$p = 0.004$.

The half-cut entropy saturates at an $L$-independent ceiling. At
$p = 0.004$ the $L = 512$ chain reaches $S = 106$ bits, close to the
predicted ceiling $0.46/p = 115$ bits. For every $p \ge 0.04$ the entropy
varies by less than $20\%$ across $L = 64$ to $512$, an eightfold increase
in size over which a volume law would grow eightfold.

\begin{figure*}[t]
\includegraphics[width=\textwidth]{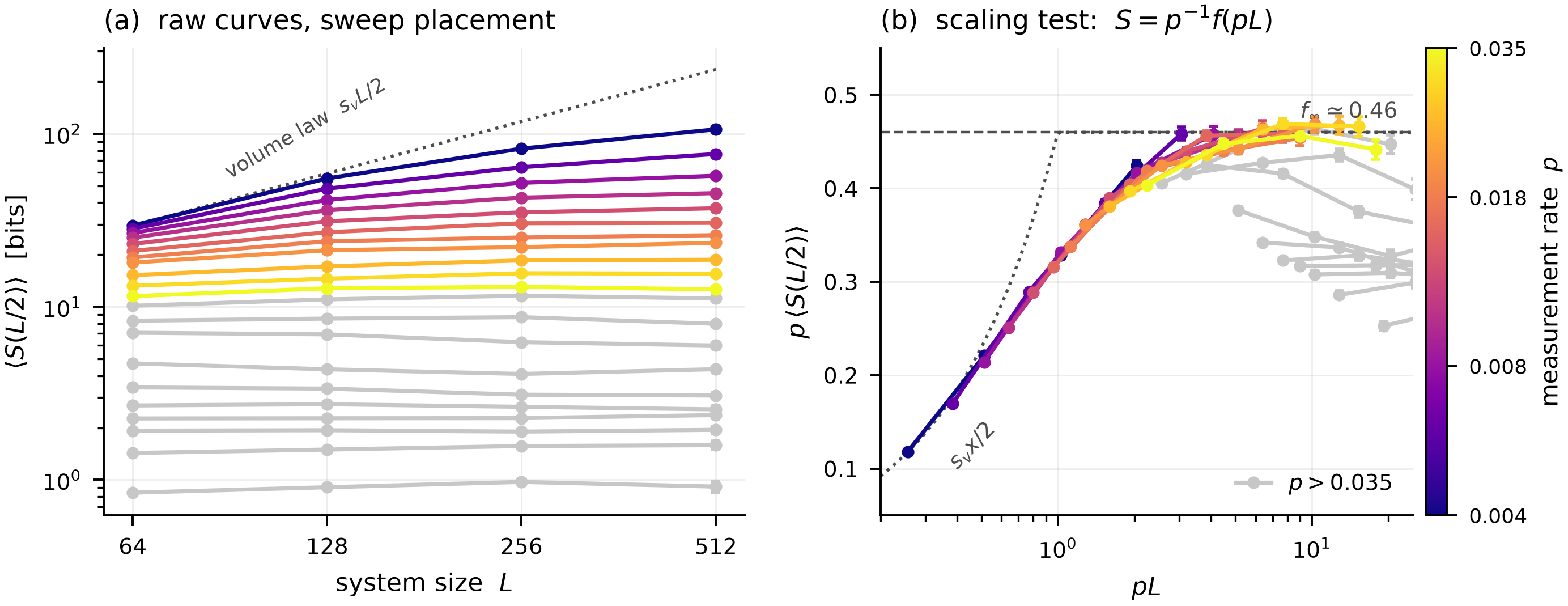}
\caption{Ballistic scaling test for sweep placement, $L = 64$--$512$.
(a) Raw half-cut entropy versus system size. Curves peel away from the
volume law $s_v L / 2$ (dotted) and flatten. (b) The same data rescaled:
$p\,S$ against $pL$ collapses onto a single curve for $p \le 0.035$
(colored), with small-argument slope $s_v/2$ and plateau
$f_\infty \simeq 0.46$ (dotted and dashed guides). Rates above $0.035$
(grey) fall below the plateau, as expected from lattice-scale corrections
once $1/p$ approaches the gate spacing.}
\label{fig:collapse}
\end{figure*}

\subsection{A ballistic argument and its scaling form}
These observations follow from a ballistic argument, which we now state
quantitatively. The sweep front advances $pL$ sites per layer on average,
so it returns to any given cut after $1/p$ layers, independent of $L$. In
the steady state, the time $\tau$ elapsed since the front last crossed a
given cut is uniformly distributed on $[0, 1/p)$. Between passages, the
entanglement across the cut regrows at most ballistically from the
remainder $S_r$ left by the passage, $S(\tau) \le S_r + v_E \tau$, where
$v_E$ is the entanglement growth velocity of the unitary dynamics.
Averaging over $\tau$ gives the steady-state bound
\begin{equation}
\langle S \rangle \;\le\; S_r + \frac{v_E}{2p},
\label{eq:ceiling}
\end{equation}
which is independent of $L$: an area law at every fixed $p > 0$, that is,
$p_c^{\rm sweep} = 0$ in the thermodynamic limit. Finite systems
interpolate between this ceiling and the volume law $S \approx s_v L/2$ of
the weakly monitored regime. Equating the two scales gives the crossover
condition $s_v L / 2 \sim f_\infty / p$, that is, $pL = O(1)$: the only
scale in the problem is $1/p$, and the scaling form
\begin{equation}
S(p, L) = \frac{1}{p}\, f(pL),
\;\;
f(x) \xrightarrow{\,x \to 0\,} \frac{s_v x}{2},
\;\;
f(x) \xrightarrow{\,x \to \infty\,} f_\infty,
\label{eq:scaling}
\end{equation}
follows. Two independent checks confirm Eq.~\eqref{eq:scaling}. First, it
collapses the data for $p \le 0.035$ across $L = 64$--$512$
(Fig.~\ref{fig:collapse}), with plateau $f_\infty \simeq 0.46$ and
small-argument slope $s_v/2 \simeq 0.46$, that is, an entropy density
$s_v \simeq 0.92$ bits per site. The latter is measured at the smallest
accessible $pL = 0.26$ and is a lower bound on the $p \to 0$ value, which
approaches the maximal density of a random stabilizer state. The two
asymptotes of $f$ meet at $pL \simeq 2 f_\infty / s_v \simeq 1$. Second,
Eq.~\eqref{eq:scaling} predicts that finite-size $\I_3$ crossings occur at
fixed $p^\ast L$, that is, $p^\ast \propto 1/L$ rather than
$p^\ast \to \mathrm{const}$ as in Eq.~\eqref{eq:fss}. The measured
crossings satisfy $p^\ast \sqrt{L_1 L_2} \approx 0.75$, an $O(1)$
constant, as predicted.

The plateau can be checked against an independent measurement. In the
unitary circuit ($p = 0$) we measure an entanglement growth velocity
$v_E = 0.647 \pm 0.005$ bits per layer at $L = 256$. If each sweep passage
reset the entanglement at a cut to zero, Eq.~\eqref{eq:ceiling} would
saturate at $S_r = 0$ and give $f_\infty = v_E/2 = 0.32$. The measured
plateau lies above this estimate. The direction of the discrepancy is
expected: the sweep measures sites rather than bonds, so a passage removes
only part of the entanglement across a cut and regrowth starts from a
nonzero remainder $S_r$. The scaling form itself, which is the substantive
claim, does not depend on this coefficient.

\begin{figure*}[t]
\includegraphics[width=\textwidth]{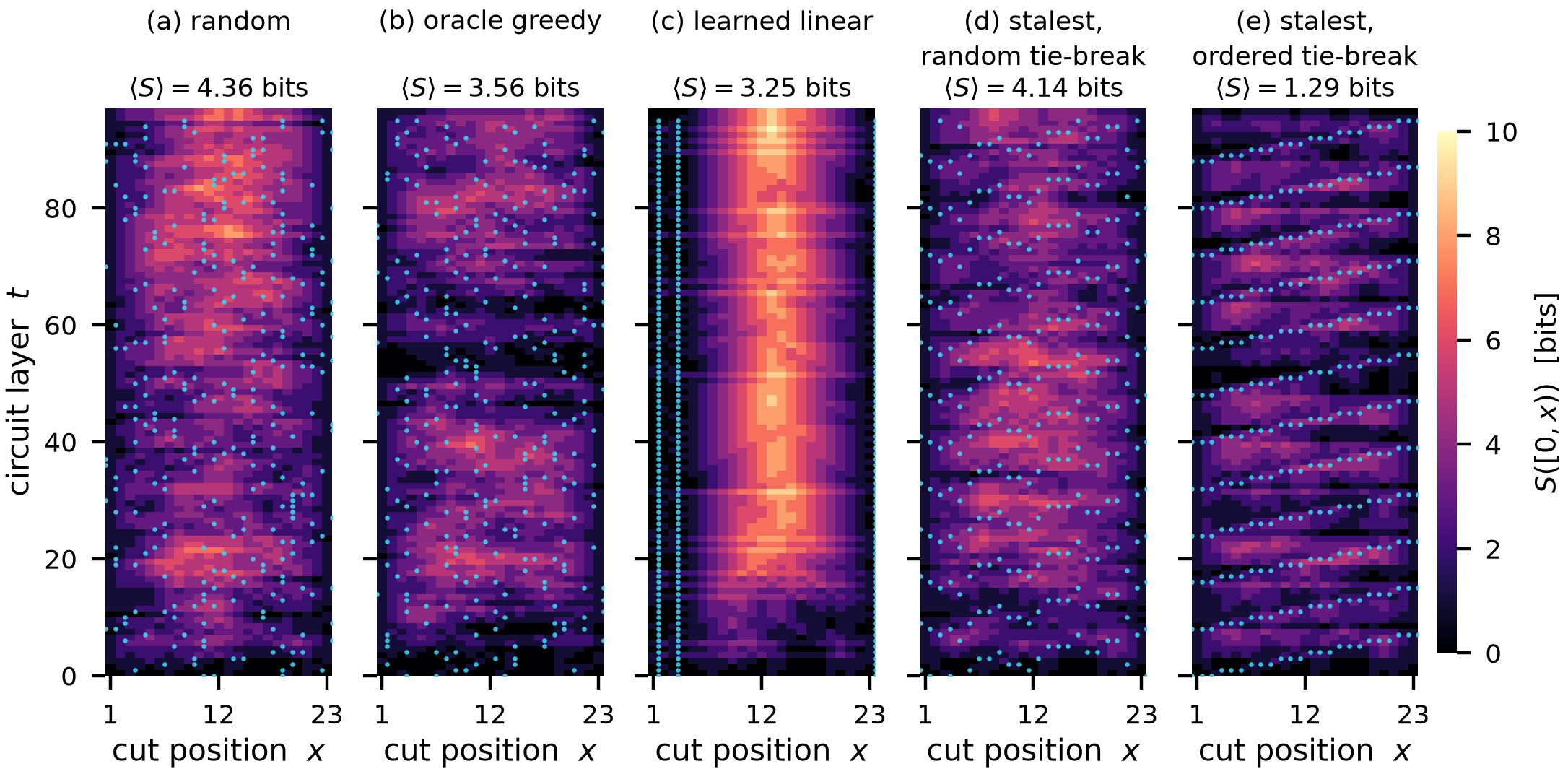}
\caption{Entanglement profile $S([0,x))$ as a function of cut position $x$
and layer $t$ under five placement policies at matched budget ($L = 24$,
$k = 3$ per layer, one trajectory each, shared color scale). Blue dots
mark measurement events; titles give the ensemble mean over $400$
episodes. Panels (d) and (e) apply the same rule, ``measure the $k$ least
recently measured sites,'' and differ only in how ties are broken: at
random in (d), in position order in (e). Random tie-breaking freezes a
scattered permutation and performs like random placement; ordered
tie-breaking produces the advancing front of the sweep and the lowest
entropy of any policy studied. The learned linear policy (c) pins
measurements at two fixed sites and fences entanglement into the
unmeasured region.}
\label{fig:spacetime}
\end{figure*}

\subsection{The effect is carried by spatial order}
\label{sec:order}
Table~\ref{tab:policies} isolates what makes the sweep effective, at
$L = 24$, $k = 3$ measurements per layer, and $T = 96$.

Coverage is not the mechanism. Shuffled coverage visits every site exactly
as often as the sweep but in a spatially unstructured order. It improves
on random placement by $6\%$, against the sweep's $70\%$.

State information is not the mechanism either. The oracle-greedy policy
has full access to the quantum state and measures the sites of largest
local entanglement. It still loses to the blind sweep by a factor of
$2.8$. Myopic use of complete quantum information is worse than a fixed
geometric pattern that uses no information at all.

The mechanism is spatial order, and the sharpest demonstration needs only
one policy. Stalest placement scores each site by the time since it was
last measured and takes the top $k$. During the first coverage period all
unmeasured sites carry the same score, so the rule determines the set of
measured sites but leaves their order free. Resolving that degeneracy at
random gives $4.14 \pm 0.06$ bits, statistically indistinguishable from
shuffled coverage. Resolving it in position order gives
$1.29 \pm 0.04$ bits. Same rule, same information, same budget, same
coverage, a factor of $3.2$ in the steady-state entropy. The reason is
visible in Fig.~\ref{fig:spacetime}: whichever order the first period lays
down is the order staleness reproduces forever after, so a single
arbitrary tie-break decides between a scattered permutation and an
advancing contiguous front. The distinction survives in $\I_3$, which is
$-0.99 \pm 0.05$ bits under random tie-breaking and identically zero in
every trajectory under ordered tie-breaking.

\begin{table}
\caption{Final half-cut entropy at matched budget
($L = 24$, $k = 3$ per layer, $T = 96$; mean $\pm$ s.e.m.\ over $400$
episodes). $H_{\rm rec}$ is the exact record entropy per measurement,
from the lemma of Sec.~\ref{sec:model}. The last two rows apply the same
staleness rule and differ only in how ties are broken.}
\label{tab:policies}
\begin{ruledtabular}
\begin{tabular}{lccc}
Policy & Information & $\langle S(L/2)\rangle$ [bits] & $H_{\rm rec}$ \\
\hline
Random placement          & none      & $4.36 \pm 0.06$ & $0.95$ \\
Shuffled coverage         & none      & $4.08 \pm 0.06$ & $0.96$ \\
Oracle greedy             & quantum   & $3.56 \pm 0.06$ & $0.99$ \\
Learned linear (CEM)      & classical & $3.26 \pm 0.09$ & $0.88$ \\[2pt]
Stalest, random tie-break & classical & $4.14 \pm 0.06$ & $0.97$ \\
Stalest, ordered tie-break \\
\quad (the sweep)         & classical & $1.29 \pm 0.04$ & $0.94$ \\
\end{tabular}
\end{ruledtabular}
\end{table}

\section{The information ledger}
\label{sec:ledger}

\begin{figure}
\includegraphics[width=\columnwidth]{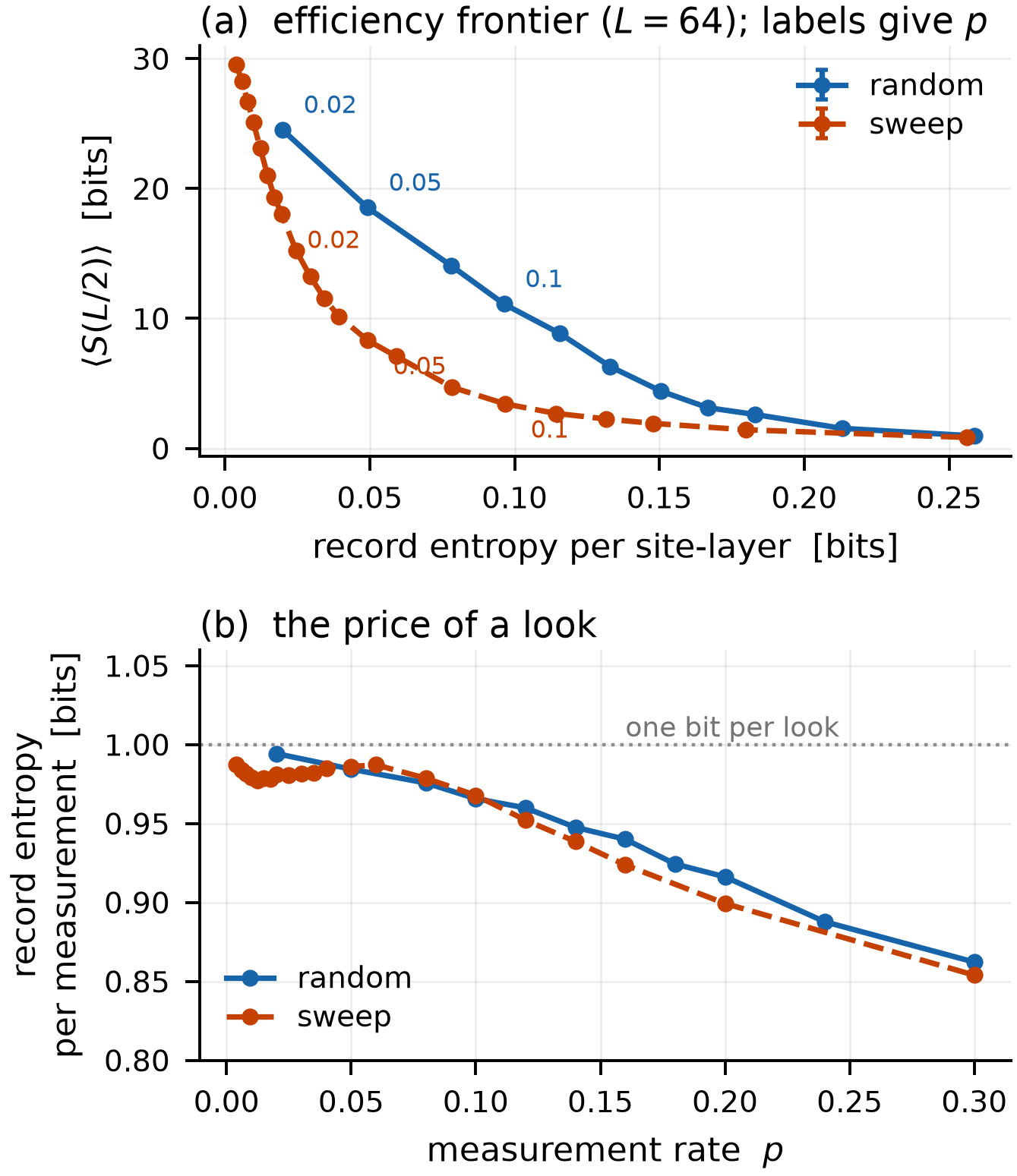}
\caption{The information ledger at $L = 64$. (a) The sweep dominates the
entropy-versus-record-cost frontier: at every budget of recorded bits it
reaches a lower entanglement entropy than random placement. Labels give
the corresponding measurement rate. (b) Both placements pay between
$0.85$ and $0.99$ bits of record entropy per measurement over the whole
range of rates, so the advantage in (a) is not cheaper information.}
\label{fig:ledger}
\end{figure}

The lemma of Sec.~\ref{sec:model} bounds what any placement policy can
achieve per recorded bit.

\emph{Proposition (one bit per bit).} In stabilizer dynamics, a
deterministic single-site measurement leaves $S_A$ unchanged for every
region $A$, and a nondeterministic one changes each $S_A$ by $0$ or $-1$.
Consequently, along any trajectory, the total entanglement removed by
measurements across any fixed cut is bounded by the record entropy:
\begin{equation}
\Delta S_{\rm meas} \;\le\; H_{\rm rec}.
\label{eq:onebit}
\end{equation}

\emph{Proof.} If the outcome is deterministic, $\pm Z_q$ is in the
stabilizer group, the projector acts as the identity on the state, and no
entropy changes. If the outcome is random, the generators can be chosen so
that exactly one anticommutes with $Z_q$, and the measurement replaces
that single generator by $\pm Z_q$. The generator matrices before and
after differ in one row, so $\mathrm{rank}_{\mathrm{GF}(2)}(G|_A)$ in
Eq.~\eqref{eq:rank} changes by at most one, and since the two outcomes
yield identical entropies, monotonicity of entanglement under local
operations forbids an increase. Summing over the measurements of a
trajectory and counting one bit per random outcome (the lemma) gives
Eq.~\eqref{eq:onebit}. \hfill$\square$

Equation~\eqref{eq:onebit} says that a demon buys at most one bit of
disentanglement per bit it must later erase. It fixes the exchange rate
but not how well a policy uses it. Figure~\ref{fig:ledger} measures that
directly, by plotting steady-state entropy against record entropy per
site-layer, parametrically in $p$, at $L = 64$.

The sweep dominates the frontier. At every record-entropy budget it
reaches a lower entanglement entropy than random placement. Read
vertically at a fixed budget of $0.05$ bits per site-layer, random
placement leaves $18.5$ bits of half-cut entropy and the sweep leaves
$8.3$.

The right panel shows what the advantage is not. Both protocols pay
between $0.85$ and $0.99$ bits of record entropy per measurement across
the full range of rates, because in and near the volume-law regime almost
every measured outcome is nondeterministic, and the two curves track each
other to within $0.03$ bits everywhere. The sweep does not buy cheaper
information. It allocates equally priced bits better.

Two further observations sharpen the point. First, the sweep is outcome
independent. Its placement decisions use no information from the record,
so the demon needs no working memory for control, and the record can be
erased as it is written at the Landauer rate without affecting the
protocol. The entire entanglement suppression of Sec.~\ref{sec:geometry}
is achieved by a demon with zero control-relevant information. That
sharpens the contrast with feedback-based adaptive circuits, where outcome
information drives the
dynamics~\cite{Iadecola2023,Sierant2023,ODea2024,Buchhold2022,Piroli2023}.
Second, the oracle-greedy policy consumes the most information of any
policy in Table~\ref{tab:policies}, $0.99$ bits per measurement, and still
loses to the sweep by a factor of $2.8$. Information about the state is
neither necessary nor sufficient for effective placement at this
granularity. Geometry is.

\section{Learned placement policies}
\label{sec:learned}

We ask whether placement structure can be learned rather than designed.
Policies observe only classical record features per site: time since the
site was last measured, time since either neighbor was last measured, last
outcome, empirical outcome bias, and layer parity. Actions select $k$
distinct sites per layer, so all comparisons remain at matched budget.

We train two policy classes to minimize the final half-cut entropy. The
first is a linear scoring policy, which takes the top $k$ sites under a
linear function of the site features, trained with the cross-entropy
method over $3{,}840$ episodes. The second is a translation-equivariant
circular convolutional network with a recurrent global memory, trained
with proximal policy optimization over $25{,}600$ episodes; actions are
sampled without replacement through a Plackett--Luce factorization, which
gives exact log-probabilities. The recurrent state exists precisely so
that the policy class can express time-advancing patterns such as sweeps.
Architecture and hyperparameters are given in Appendix~\ref{app:learning}.

Table~\ref{tab:learned} reports matched-budget evaluations. Neither
learner finds the sweep, and the two fail in the same direction.

The CEM linear policy improves on random placement by $25\%$ at $L = 24$
(Table~\ref{tab:policies}) and converges to a fencing strategy: it pins
measurements at a small set of fixed sites and quarantines entanglement
between them (Fig.~\ref{fig:spacetime}c). Fencing is locally sensible and
globally suboptimal. A fence is a static cut, and the region it protects
keeps whatever entanglement it has.

The PPO policy does not improve on random placement at all
($5.38 \pm 0.08$ against $5.27 \pm 0.08$), despite a policy class that
contains moving patterns and $6.7$ times more training episodes than the
CEM run. Its training curve is flat over the final $300$ updates. Its
stochasticity is doing the work: decoding the same policy greedily
performs worse ($5.87 \pm 0.10$) and writes the fewest random bits of any
policy in the table, $0.87$ per measurement, which is the signature of
collapse onto repeatedly measured sites. That is fencing again. The sweep
outperforms every learned and informed policy by more than a factor of
two.

\begin{table}
\caption{Final half-cut entropy at matched budget
($L = 32$, $k = 4$ per layer, $T = 128$; mean $\pm$ s.e.m.\ over $300$
episodes). $H_{\rm rec}$ is the exact record entropy per measurement.}
\label{tab:learned}
\begin{ruledtabular}
\begin{tabular}{lccc}
Policy & Information & $\langle S(L/2)\rangle$ [bits] & $H_{\rm rec}$ \\
\hline
Random placement          & none      & $5.27 \pm 0.08$ & $0.96$ \\
Shuffled coverage         & none      & $4.94 \pm 0.08$ & $0.97$ \\
Stalest, random tie-break & classical & $4.87 \pm 0.08$ & $0.97$ \\
PPO (stochastic)          & classical & $5.38 \pm 0.08$ & $0.96$ \\
PPO (greedy decoding)     & classical & $5.87 \pm 0.10$ & $0.87$ \\
Oracle greedy             & quantum   & $4.24 \pm 0.08$ & $0.99$ \\
Sweep                     & none      & $1.93 \pm 0.06$ & $0.95$ \\
\end{tabular}
\end{ruledtabular}
\end{table}

\subsection{Why the learners miss it}
Section~\ref{sec:order} localizes the difficulty. The optimal policy and a
policy no better than random differ only in how a degenerate score is
resolved. Both learners are score-based: they compute a real-valued
logit per site and take the top $k$. Such a parameterization controls
which sites are selected, and it controls that smoothly. It does not
parameterize the order in which equal scores are resolved. In our
implementations that order comes from a random perturbation, so the
learners search a family whose best member is stalest placement with
random tie-breaking, worth $4.87 \pm 0.08$ bits at $L = 32$, and not the
sweep, worth $1.93 \pm 0.06$.

Symmetry makes this structural rather than incidental. The PPO network is
translation equivariant on a ring, and the initial state is translation
invariant, so its logits are exactly degenerate on the first layer. No
setting of the weights breaks that degeneracy; only the sampling does, and
sampling breaks it at random. A policy that could condition on absolute
position would escape this, at the cost of the equivariance that makes the
architecture efficient.

Myopia compounds the problem. Moving one measurement by one site changes
the final entropy very little, so the reward landscape around unstructured
policies is nearly flat, while the sweep's advantage is a property of a
long temporally coherent trajectory of placements. Local policy
improvement does not assemble such trajectories from unstructured starting
points, and both learners instead settle into the nearest attractor,
fencing, which is a static approximation to cutting.

Learning where to measure is therefore hard for a specific and diagnosable
reason. The good strategy is not complex, and it is not expensive to
execute. It lives in a discrete ordering degree of freedom that
score-based policy classes do not expose to the optimizer, and its value
is invisible to myopic credit assignment even when it is exposed.

\section{Discussion}
\label{sec:discussion}

The phase diagram of monitored dynamics is a property of the placement
process, not only of the measurement rate. The existing
structured-placement literature, in which quasiperiodic
sublattices~\cite{Shkolnik2023} and quenched disorder~\cite{Zabalo2023}
modify the universality class but preserve the transition, might suggest
that placement structure is a marginal perturbation. The sweep shows the
opposite. Structure that advances in time removes the volume-law phase
entirely at matched budget.

The distinction between static and time-advancing structure is therefore
qualitative. A static pattern leaves unmeasured regions where entanglement
survives indefinitely, which is exactly the fencing failure mode both of
our learners find. A sweep guarantees that every site is refreshed on the
$L$-independent timescale $1/p$, and ballistic regrowth cannot outrun it.

Three implications follow. For theory, the measurement rate alone is an
incomplete control parameter, and statements about the volume-law phase
implicitly assume unstructured placement. For classical simulation, a
sweeping measurement schedule keeps monitored Clifford dynamics in a
low-entanglement regime at rates far below the random-placement
transition, which may extend to tensor-network simulability of
non-Clifford monitored systems. For experiments in which mid-circuit
measurements are slow or costly, entanglement suppression per measurement
is the relevant figure of merit, and placement geometry delivers more of
it than either coverage or state knowledge.

Several questions remain open. First, the crossover function $f$ and its
universality: whether $f$ depends on the gate ensemble only through $v_E$,
and how the sweep front's partial cutting renormalizes the plateau.
Second, dimensionality: in two dimensions a sweeping front of measured
sites is a line, and whether an analogous single-scale collapse holds is
unknown. Third, optimality: among outcome-independent placement processes
at fixed budget, whether the sweep minimizes the steady-state entropy, and
what the corresponding bound is. Fourth, learning: our diagnosis suggests
concrete remedies, including policy classes that parameterize an ordering
directly, curricula that start near sweep-like policies, and objectives
with longer-horizon credit assignment. Whether any of them reaches the
geometric optimum bears on the broader question of when reinforcement
learning can discover temporally coherent control strategies. Finally, the
ledger beyond stabilizer circuits: for non-Clifford monitored dynamics the
record entropy is no longer a simple count, and the tradeoff between
entanglement steered and information recorded becomes a genuine
thermodynamic optimization.

\begin{acknowledgments}
Simulations were performed in part on the HYAK computing cluster at the
University of Washington. AI-based tools were used to assist with
literature search, code development, data analysis, and manuscript
preparation; the author directed this work and verified all results.
\end{acknowledgments}

\section*{Data and code availability}
The simulation code, the trained policy checkpoints, and the numerical
data behind every figure and table are available from the author on
request.

\appendix

\section{Numerical methods}
\label{app:numerics}

\emph{Circuits.} Chains of $L$ qubits with periodic boundary conditions,
$L$ divisible by 4. Brickwork layers apply independent uniformly random
two-qubit Clifford gates to even bonds on even layers and odd bonds
(including the wrap bond) on odd layers. Projective $Z$ measurements
follow each gate layer according to the placement process. All runs use
depth $T = 4L$ from the product initial state $|0\rangle^{\otimes L}$.

\emph{Entropies.} All entropies use Eq.~\eqref{eq:rank} on the canonical
stabilizer generators. $\I_3$ uses Eq.~\eqref{eq:i3} with contiguous
quarters $A, B, C, D$.

\emph{Crossings.} We take the difference
$\Delta(p) = \I_3^{(L_1)}(p) - \I_3^{(L_2)}(p)$ on the shared $p$ grid. A
sign change of $\Delta$ between adjacent grid points counts as a crossing
only if $|\Delta|$ exceeds twice the combined standard error at both
flanking grid points. This local criterion rejects the spurious sign
changes that occur deep in the area-law regime, where both curves vanish
within noise. Crossing positions are obtained by linear interpolation.
Under this criterion four of the six size pairs cross significantly for
random placement, giving $p_c = 0.158 \pm 0.003$, and two do for the
sweep. A weaker criterion that requires significance only somewhere on
each side of the sign change admits all six random-placement pairs plus
one spurious crossing at $p = 0.179$, and gives $p_c = 0.162 \pm 0.008$;
the two criteria agree within their quoted spreads.

\emph{Statistics.} Trajectory counts per $(L, p)$ point are $2000$, $1000$,
$400$, and $128$ for $L = 64$, $128$, $256$, and $512$. Quoted
uncertainties on entropies are standard errors over trajectories. The
uncertainty on $p_c$ is the spread of the significant pairwise crossings,
not a fit error, and with four crossings it is a coarse estimate. The
random-placement ledger data of Fig.~\ref{fig:ledger} use $100$
trajectories per point. The evidence that the sweep crossings recede rests
on two significant crossings and on the absence of a significant crossing
for every pair involving $L = 256$ or $512$; the collapse of
Fig.~\ref{fig:collapse}, which uses all $84$ $(L,p)$ cells, is the
stronger of the two tests of Eq.~\eqref{eq:scaling}.

\emph{Velocity.} $v_E$ is the slope of the ensemble-averaged $S(L/2, t)$
at $p = 0$, $L = 256$, fit over the central $80\%$ of a depth-$0.6L$
window. The quoted uncertainty is the standard error of the mean over the
$16$ per-trajectory slopes.

\section{Learned policy details}
\label{app:learning}

\emph{Features.} Each site presents five classical features: time since
the site was last measured and time since either neighbor was last
measured (both normalized by $L$), the last outcome ($\pm 1$, or 0 if
never measured), the empirical fraction of $+1$ outcomes at the site, and
the layer parity.

\emph{CEM linear policy.} Scores are a linear function of the site
features plus a global mean-feature term (10 parameters); the $k$ highest
scores are measured, with ties broken by an infinitesimal random
perturbation. Training used the cross-entropy method with population $24$,
elite fraction $1/4$, additive noise floor $0.05$, and $10$ iterations of
$16$ episodes per candidate, for $3{,}840$ training episodes in total.
Reported evaluations use $400$ fresh episodes.

\emph{PPO policy.} Two circular convolutions (kernel 3, 32 channels,
ReLU) produce per-site embeddings; their mean feeds a GRU cell (32 hidden
units) whose state is broadcast back to every site; a $1 \times 1$
convolution over the concatenation yields per-site logits $\ell_i$, and a
linear head on the pooled representation yields the value estimate. The
network has no absolute-position input, so it is exactly equivariant under
rotations of the ring. Actions are $k$ distinct sites sampled without
replacement by sequential softmax (Plackett--Luce),
\begin{equation}
\pi_\theta(a_1, \ldots, a_k \mid s)
 = \prod_{j=1}^{k}
   \frac{e^{\ell_{a_j}}}
        {\sum_{i \notin \{a_1, \ldots, a_{j-1}\}} e^{\ell_i}},
\label{eq:pl}
\end{equation}
whose log-probability is exact, as required by the PPO objective
\begin{equation}
\mathcal{L}(\theta) = \mathbb{E}_t\!\left[
 \min\!\big(r_t(\theta) \hat A_t,\;
 \mathrm{clip}(r_t(\theta), 1{-}\epsilon, 1{+}\epsilon)\, \hat A_t\big)
\right],
\label{eq:ppo}
\end{equation}
where $r_t(\theta) = \pi_\theta(a_t|s_t)/\pi_{\theta_{\rm old}}(a_t|s_t)$
and $\hat A_t$ is the generalized advantage estimate. Training: Adam with
learning rate $3 \times 10^{-4}$, discount $\gamma = 0.995$, GAE
$\lambda = 0.95$, clip $\epsilon = 0.2$, entropy bonus $0.01$, value
coefficient $0.5$, 4 epochs per update, 64 episodes per update, 400
updates, for $25{,}600$ training episodes in total at $L = 32$, $k = 4$,
$T = 128$. The reward is the per-layer decrease of the half-cut entropy,
plus the negative final half-cut entropy at the terminal step. Reward
shaping uses quantum-side information during training only; the deployed
policy acts on classical features alone. Reported evaluations use $300$
fresh episodes.

\bibliography{refs}

\end{document}